\documentstyle[aps]{revtex}

\begin{document}
\draft
\title{ The Magnus force in superfluids and superconductors}
\author{E.B. Sonin}

\address{Low Temperature Laboratory, Helsinki University of Technology, 02150
Espoo, Finland \\ and \\ Ioffe Physical Technical Institute, St.~Petersburg
194021, Russia}

\date{\today} \maketitle

\begin{abstract}
The forces on the vortex, transverse to its velocity, are
considered. In addition to the superfluid Magnus force from the condensate
(superfluid component), there are transverse forces from thermal
quasiparticles and external fields violating the translational invariance. The
forces between quasiparticles and the vortex originate from interference of
quasiparticles with trajectories on the left and on the right from the vortex
like similar forces for electrons interacting with the thin magnetic-flux tube
(the Aharonov-Bohm effect). These forces are derived in the Born approximation
for phonons from the equations of superfluid hydrodynamics, and
for BCS quasiparticles from the Bogolyubov-de Gennes equations.

The effect of external fields violating translational invariance is analyzed
for vortices in the two-dimensional Josephson junction array. The symmetry
analysis of the classical equations for the array shows that the total
transverse force on the vortex vanishes. Therefore the Hall effect which is
linear in the transverse force is absent also. This means that the
Magnus force from the superfluid component {\em exactly} cancels with the
transverse force from the external fields.

The results of other approaches are also brought together for discussion.
\end{abstract}


\section{Introduction} \label{In}

The Magnus force on a vortex has long been known in classical
hydrodynamics \cite{Magnus}. This force appears if the vortex moves with
respect to a liquid. The force is normal to the relative vortex velocity and
therefore is reactive and does not produce a work. In general, such  a force
arises always when a body with a flow circulation around it  moves through a
liquid or a gas (the Kutta-Joukowski theorem). The most important example is
the lifting force on a wing of a aeroplane which keeps the aeroplane in
the air.

The key role of the Magnus force in vortex dynamics has become clear from the
very beginning of studying superfluid hydrodynamics \cite{HV,Hall}. The
superfluid Magnus force was defined as   a force between a vortex and a
superfluid and therefore was proportional to the superfluid density $\rho_s$.
But in the two-fluid hydrodynamics the superfluid Magnus force is not the only
force on the vortex transverse to its velocity: there was also a transverse
force between the vortex and quasiparticles moving with respect to the vortex.
The transverse force from rotons was found by Lifshitz and Pitaevskii
\cite{LP} from the quasiclassical scattering theory. Later  Iordanskii
\cite{I} revealed the transverse force from phonons which was equal in
magnitude  and opposite in sign with the quasiclassical force of Lifshitz and
Pitaevskii. From the very beginning the Iordanskii force was a controversial
matter. Iordanskii suggested that his force and the  Lifshitz-Pitaevskii
force  were of different origins and for rotons they should be summed. As a
result, he concluded that the transverse force from rotons vanished. But the
analysis done in Ref. \cite{S} demonstrated that the Iordanskii force for
rotons is identical to the Lifshitz-Pitaevskii force and they must not be
added.  In addition, the Lifshitz-Pitaevskii force from rotons was calculated
in the original paper \cite{LP} with a wrong sign. After its correction the
transverse force on the vortex had a same sign and a value both for rotons
(the Lifshitz-Pitaevskii force) and for phonons (the Iordanskii force). In
the same paper \cite{S} it was pointed out that the Iordanskii force for
phonons and rotons results from interference between quasiparticles which
move past the vortex on the left and on the right sides with different phase
shifts, like in the Aharanov-Bohm effect \cite{AB}.

In the theory of superconductivity the Magnus force appeared first in the
paper by Nozi\`eres and Vinen \cite{NV}. In clean superconductors  the  BCS
quasiparticles produce an additional transverse force on the vortex
\cite{GS,KK-1} analogous to the Iordanskii force in superfluids. The total
transverse force is responsible for the Hall effect in the mixed state of a
superconductor. But the Hall effect was rather weak in classical
superconductors. An explanation of it was suggested by Kopnin and Kravtsov
\cite{KK} (see also Ref. \cite{KL}): impurities interact with quasiparticles
bound in the vortex core and this interaction produces an additional
transverse force on the vortex. In contrast with the quasiparticle transverse
force which increases the total transverse force, the impurity force
decreases it and in a dirty superconductor completely cancels the Magnus
force. As a result, the strong Hall effect is possible only in superclean
superconductors. The transverse force from the bound states in the core has
been recently rephrased in terms of the spectral flow through the
quasiparticle bound states \cite{KVP}.

A new wave of interest to the Magnus force came with discovery of high-T$_c$
superconductivity. A few reasons of this interest might be mentioned: (i) The
so-called Hall anomaly was observed \cite{HA}: near $T_c$ the sign of the
Hall voltage is opposite to that expected from the standard vortex dynamics.
(ii) It has become possible to obtain superclean single crystals with large
Hall angle as predicted in the superclean limit of the theory of Kopnin and
Kravtsov. This made possible to observe magnetoresonances in the a.c.
response of the superconducting single crystals connected with waves
propagating along vortices \cite{Ong,LKST}. (iii) The effective Magnus force
governs  quantum vortex  nucleation in clean high-T$_c$ superconductors
intensively discussed now \cite{Bl,SQ}.

Despite a lot of work done to understand and calculate the Magnus force, it
still remains to be a matter of controversy with a number conflicting points
of view on it. A new discussion has been launched by the paper of Ao and
Thouless \cite{AT}. The main claim of Ao and Thouless is that there is an
universal {\em exact} expression for the total transverse  force on the
vortex (the  effective Magnus force) which does not depend on the presence of
quasiparticles  or impurities. This force derived from the concept of the
geometrical phase (the Berry phase) coincides with the superfluid Magnus force
and therefore is proportional to the superfluid density. According to Ao and
Thouless, there is no transverse force on the vortex from quasiparticles and
impurities, though they might influence the value of the superfluid density
and thereby influence the amplitude of the Magnus force.

The Ao-Thouless theory is in an evident disagreement with the previous
calculations of the transverse force on the vortex (the effective Magnus
force) in superfluids and superconductors reviewed above. It attracted a
great attention and has been supported in a number of recent publications of
other theorists (see e.g. Refs. \cite{Ga,Sim,GaS}). If the Ao-Thouless theory
were true, it would be necessary to revise the whole basis of the vortex
dynamics. For example, on the basis of the Ao-Thouless theory \v{S}imanek
\cite{Sim} suggested that quantum vortex tunnelling is governed by the Magnus
force obtained from the Berry phase approach, i.e., proportional to the
superfluid density, in contradiction to the previous theory \cite{Bl,SQ}.
Therefore it is important to understand what the Magnus force is and whether
the Ao-Thouless theory is true or not.

The present paper is to analyze this controversy. Among the sources of
controversy there is semantics. Therefore it is important to define force
terminology from the very beginning. The word {\em force} itself is only a
label to describe a transfer of momentum between two objects. Before using
these labels one must to analyze the momentum balance for any object and only
then to label various contributions to these balances as forces. Keeping this
in mind, the forces under discussions may be defined as the following:
\begin{itemize}
\item
There is a momentum transfer between a superfluid and a vortex. It is
revealed analyzing the momentum balance for the superfluid moving with the
velocity $\vec v_s$ whereas the vortex moves  through the superfluid with a
different velocity $\vec v_L$. This momentum transfer is the
\underline{superfluid Magnus force}. It is proportional to the superfluid
density $\rho_s$ and transverse to the relative velocity $\vec v_L - \vec
v_s$.
\item
Analyzing the momentum balance for the {\em whole} translationally invariant
liquid (including the superfluid and normal parts of it) around the vortex
one may reveal a contribution presenting the momentum transfer between the
vortex moving with the velocity $\vec v_L$ and the normal fluid (the gas of
quasiparticles) moving with the velocity $\vec v_n$. This force is
proportional to the relative velocity $\vec v_L - \vec v_n$ and has the
components longitudinal and transverse to $\vec v_L - \vec v_n$. The
transverse component is the \underline{Iordanskii force}.
\item
If there is no translational invariance, as in a dirty superconductor, the
momentum balance for the whole liquid must include also forces external for
the liquid, namely, the momentum transfer to the impurities rigidly connected
with the crystal lattice. When the latter is at rest, this momentum transfer,
or the force from impurities, is proportional to the absolute value of the
vortex velocity $\vec v_L$. Its component transverse to $\vec v_L$ is the
\underline{Kopnin-Kravtsov force}.
\item
The momentum balance for the whole liquid around the vortex is at the same
time an equation from which one must find the vortex velocity $\vec v_L$.
Therefore it is useful to collect all the terms proportional to $\vec v_L$
together. After it the term uniting all contributions transverse  to $\vec
v_L$ is the total transverse force on the vortex, or the \underline{effective
Magnus force}.
\end{itemize}

Thus in general three forces contribute to the effective Magnus force: the
superfluid Magnus force, the Iordanskii force, and the Kopnin-Kravtsov force.
Experimentally they can determine the effective Magnus force, but not  the
``bare'' superfluid Magnus force. In rotating superfluids the effective Magnus
force determines the mutual friction.  One can find the latest reviews of the
experiment and the theory on mutual friction in $^3$He in Refs. \cite{JH,JK}.
In superconductors the Hall conductivity \cite{HA,HK} and the acoustic Faraday
effect for the transverse ultrasound wave propagating along vortices \cite{US}
are linear in the effective Magnus force. The process of vortex quantum
tunnelling is also influenced by the effective, but not the superfluid Magnus
force.  So the final outcome of the theory must be the amplitude of the
effective Magnus force. Its presentation as a combination of three forces is
an intermediate stage of the theory. In fact, this presentation is valid only
if (i) the number of quasiparticles is not too large and their mutual
interaction is not strong; (ii) external fields violating translational
invariance are not too strong. The first condition is violated close to
$T_c$, where other approaches based on the Ginzburg-Landau theory (or its
analogue for superfluids, the Ginzburg-Pitaevskii theory) must be used
\cite{RMP,16}. The second condition doesn't hold in the Josephson junction
array considered in the present paper (see below). In these cases the theory
deals directly with the effective Magnus force in the equation of vortex
motion: its decomposition on the ``bare'' superfluid Magnus force and the
forces from quasiparticles or impurities becomes conventional and of a little
physical sense.

Whereas there is  a consensus among theorists on the superfluid Magnus force,
the Ao-Thoules theory rejects the Iordanskii force from quasiparticles and the
Kopnin-Kravtsov force from impurities claiming that amplitudes of the
effective and the superfluid Magnus forces are exactly equal. Therefore the
present paper considers the effect on the force balance from (i)
quasiparticles, and (ii) from external fields violating the translational
invariance of the superfluid.

In my analysis of the quasiparticle effect I chose the phonon-vortex
interaction which may be described by the nonlinear Schr\"odinger equation
long ago suggested for a weakly nonideal Bose-gas (the Gross-Pitaevskii
theory \cite{GP}). It is well known that the nonlinear Schr\"odinger equation
yields the usual superfluid hydrodynamics. It is a good starting point for
further discussion, which one may expect  a consensus of all parties on.
Indeed, according to Aitchison, Ao, Thouless, and Zhu \cite{AAT}, the
nonlinear Schr\"odinger equation, and the superfluid hydrodynamics derived
from it, is a reliable model to describe both Bose- and Fermi-superfluids
near $T=0$. The next step is to analyze scattering of the sound wave (phonon)
by the vortex in hydrodynamics. Just at this stage a disagreement appears. Ao
and Thouless believe that this scattering can produce only a dissipative
force on the vortex, but not a transverse one. Recently Demircan, Ao, and
Niu  \cite{DAN} tried to prove it using the Born approximation. But they
ignored peculiarities of the phonon Born scattering at small angles which
resulted in the Iordanskii force. It is important to note that the
controversy arises not from a difference in ideology: anyone is free to
choose a language to derive the Magnus force at $T=0$: either the standard
hydrodynamics, or newest topological concepts of the geometrical phase. But
there is a disagreement in calculation of integrals describing the phonon
scattering in the first order of the perturbation theory.  We hope to show in
this paper at which point  Demircan, Ao, and Niu  \cite{DAN} missed to take
into account the Aharonov-Bohm interference of phonons which was ignored by
the Ao-Thouless theory.

Now discussions around the transverse force on the vortex in the presence of
impurities violating the translational invariance concentrate mostly on the
contribution of the core bound states in Fermi superfluids. This requires a
rather  sophisticated analysis (see \cite{KVP} and references therein). In the
present paper I chose another example when the translational invariance is
absent: the two-dimensional Josephson junction array(JJA). This is a regular
lattice of nodes with the Josephson coupling   between them. Experimentally,
any  node corresponds to a superconducting island  in an artificially
prepared JJA, or to a grain in a granular superconducting  film. The behavior
of the JJA in an external magnetic field is usually described in the picture
of moving vortices similar to the mixed state of type  II superconductors.
The dynamics of vortices in JJA attracts a   great interest of
experimentalists \cite{1,2} and theorists \cite{3,4,5,6,7,9,10,11,11.1,qJJA}.
There is an intrinsic pinning of vortices at the JJA cells, and vortices can
move only if the driving supercurrent is more than the critical value. But
when they start to move, in many cases a good approximation is to replace the
lattice by a  continuous superconducting film. However, the hydrodynamic
derivation of the Magnus force is not valid since it assumes the momentum
conservation law and the translational invariance. In the  present paper it
will be shown that the Hall effect is {\em exactly} absent in the classical
theory of JJA which neglects the charge quantization. Since the Hall effect
is linear in the amplitude of the effective Magnus force, the latter also
vanishes in the classical JJA.  This statement directly follows from the
symmetry of the dynamic equations. At the same time  the superfluid density
is finite in the continuum limit of JJA and therefore the superfluid Magnus
force doesn't vanish. Therefore the theory based on the Berry phase approach
\cite{GaS}  predicted a finite effective Magnus force and the Hall effect for
JJA. Our result might be interpreted as that the superfluid Magnus force is
compensated by some force external for the liquid, like the Kopnin-Kravtsov
force in a dirty superconductor. But as mentioned above, JJA is a system with
a strong violation of translational invariance, for which this interpretation
is purely formal. Only the resultant effective Magnus force has a physical
meaning.

We start from Sec. \ref{Gen} which shows how the Magnus force appears in the
phenomenological theory of neutral and charged  superfluids. The force
terminology is also introduced explaining to which term and in which equation
any force under discussion corresponds. Section \ref{Iord} is devoted to the
transverse force between quasiparticles and the vortex (the Iordanskii force)
and its connection with the Aharonov-Bohm effect. Section \ref{Schrod} reminds
connection between the nonlinear Schr\"odinger equation for the condensate
(the Gross-Pitaevskii theory) and superfluid hydrodynamics and phonons.
Scattering of the sound wave (phonon) by the vortex in hydrodynamics is
analyzed in Sec \ref{Born} using the Born approximation and the effective
cross-sections. It is shown that the standard scattering-theory approach
fails to reveal the transverse Iordanskii force because of the divergence of
the scattering amplitude at small angles of scattering. The analysis of the
small-angle scattering is presented in Sec. \ref{sm-angle}. It reveals the
interference between quasiparticles  with trajectories on the left and on the
right from the vortex. In Sec. \ref{AB-PW} the same results is rederived
using the partial-wave expansion, and the analogy with the Aharonov-Bohm
effect is shown. A more general quasiclassical derivation of the transverse
force from the quasiparticles with an arbitrary spectrum is presented in Sec.
\ref{quasicl}. In Sec. \ref{BCS} the transverse force between the BCS
quasiparticles and the vortex is derived using the Born approximation for the
Bogolyubov-de Gennes equations. Section \ref{JJA} presents the symmetry
analysis of the classical dynamical equations for JJA which shows that the
total transverse force on the vortex in JJA vanishes and as a result of it
the Hall effect is possible only in the quantum theory of JJA which takes
into account charge quantization. The last section \ref{DS} contains the
summary and the discussion of other approaches to the problem.

\section{Where and how the Magnus force appears} \label{Gen}

\subsection{The Magnus force in classical hydrodynamics}
\label{clasMag}

As mentioned in Introduction, the Ao-Thouless theory  connects the Magnus
force with the concept of the geometrical phase.  But for better
understanding of the Magnus force origin it is worth to remind how the Magnus
force arises in classical hydrodynamics.

Let us consider an isolated straight vortex line in an incompressible inviscid
liquid. The line along the axis $z$ induces the velocity field
\begin{equation}
\vec v_v(\vec r)=\frac{\vec \kappa \times \vec r}{2\pi r^2}~.
   \label{v-v} \end{equation}
Here $\vec r$ is the position vector in the plane $xy$, and $\vec \kappa$ is
the circulation vector directed along the axis $z$. The circulation,
given by
\begin{equation}
     \kappa = \oint \vec v_v \cdot d\vec l~,
       \label{circ} \end{equation}
may have arbitrary values in classical hydrodynamics. In addition, there is a
fluid current past the vortex line with a velocity $\vec v_0$. Then the net
velocity field around the line is
\begin{equation}
\vec v(\vec r) = \vec v_v(\vec r) + \vec v_0~.
         \label{vel} \end{equation}
The Euler equation for the liquid is
\begin{equation}
\frac{\partial \vec v}{\partial t} + (\vec v \cdot \vec \nabla)\vec v= - {1
\over \rho} \vec \nabla P + {\vec F \over \rho}\delta_2(\vec r)~.
   \label{Euler} \end{equation}
Here $\rho$ is the liquid density and $P$ is the pressure. This equation
suggests that an external  $\delta$-function  force $\vec F$ is applied to the
liquid at the vortex line.

Assuming that the vortex line moves with the constant velocity $\vec v_L$, one
obtains that
\begin{equation}
\frac{\partial \vec v}{\partial t}= - (\vec v_L \cdot \vec \nabla)\vec v~.
       \label{time-der} \end{equation}
Then the Euler equation (\ref{Euler}) yields the Bernoulli law for the
pressure:
\begin{equation}
P= P_0 -{1 \over 2} \rho [\vec v(\vec r) - \vec v_L]^2 =P_0'
-{1 \over 2} \rho \vec v_v(\vec r)^2 -\rho \vec v_v(\vec r) \cdot (\vec v_0 -
\vec v_L)~.
              \label{Bernu} \end{equation}
Here $P_0$ and $P_0'=P_0 -{1 \over 2} \rho (\vec v_0 -\vec v_L)^2$ are
constants which are of no importance for the following derivation.

Next one should consider the momentum balance for a cylindrical region of a
radius $r_0$ around the vortex line. The momentum-flux tensor is given by
\cite{LL}
\begin{equation}
\Pi_{ij} =P\delta_{ij} + \rho  v_i(\vec r) v_j(\vec r)~,
    \label{mom-flux} \end{equation}
or in the reference frame moving with the vortex velocity $\vec v_L$:
\begin{equation}
\Pi'_{ij} =P\delta_{ij} + \rho  ( v_i-  v_{Li})  (v_j -v_{Lj})~.
    \label{mom-flux-L} \end{equation}

The momentum conservation law requires that the external force $\vec F$ on
the vortex line must be equal to the momentum flux through the entire
cylindrical boundary in the reference frame moving with the vortex velocity
$\vec v_L$. The latter is given by  the integral $\int dS_j \Pi'_{ij}$ where
$dS_j$ are the components of the vector $d\vec S$ directed along the outer
normal to the boundary of the cylindrical region and equal to the elementary
area of the boundary in magnitude. Then using Eqs. (\ref{v-v}),
(\ref{Bernu}), and (\ref{mom-flux-L}), the momentum balance yields the
following relation:
\begin{equation}
\rho [(\vec v_L - \vec v_0) \times \vec \kappa] =\vec F~.
   \label{Magnus} \end{equation}

On the left-hand side of this equation one can see the Magnus force as  it
comes in the classical hydrodynamics. A half of this force is due to the
Bernoulli contribution to the pressure, Eq. (\ref{Bernu}); another half is
due to the convection term $\propto v_iv_j $ in the momentum flux. The Magnus
force balances the resultant of all external forces applied to the liquid at
the vortex line (the force $\vec F$). In the absence of external forces the
vortex moves with the velocity of the liquid: $\vec v_L=\vec v_0$ (the
Helmholtz theorem).

This derivation demonstrates the classical origin of the Magnus force:
quantization of circulation is  not necessary for its existence. During the
derivation we referred to the hydrodynamic equations only at large distance
$r_0 \gg r_c$ from the vortex line. It might seem as if the fluid in the
vortex core did not matter at all. However, the derivation is based on the
assumption that the momentum is a well-defined {\em conserved} quantity
everywhere even inside of the vortex core where the hydrodynamic theory does
not hold.

\subsection{The superfluid Magnus force}
\label{supMag}

In the superfluid hydrodynamics one can refer this derivation to the
superfluid component with density $\rho_s$. The Euler equation for the
superfluid component \cite{LL} after adding the external $\delta$-function
force $\vec F$ applied at the vortex line is
\begin{equation}
\frac{\partial \vec v_s}{\partial t}+ (\vec v_s \cdot \vec \nabla)\vec v_s
= - \vec \nabla \mu  + {\vec F \over \rho_s} \delta_2(\vec r)   ~,
   \label{Euler-s0} \end{equation}
where $\mu$ is the chemical potential.

For charged superfluids (superconductors) the Euler equation should include
also the  electromagnetic forces. In particular, the chemical potential must
be replaced by the electrochemical potential. But except for vortex cores  one
may use the quasineutrality condition that the total electron charge is
approximately equal to the background ion charge. This allows to neglect the
chemical potential gradient. Finally the Euler equation may be written as
\begin{equation}
\frac{\partial \vec v_s}{\partial t}+ (\vec v_s \cdot \vec
\nabla)\vec v_s = {e \over m} \left( \vec E + {1 \over
c} [\vec v_s \times \vec H ]\right)
+ {\vec F \over \rho_s} \delta_2(\vec r)   ~,
   \label{Euler-sup0} \end{equation}
where $\vec E$ and $\vec H$ are the electric and the magnetic fields.

Let us consider a vortex line in a neutral or charged superfluids with the
velocity field Eq. (\ref{v-v}). Now circulation is quantized, and the
circulation quantum is $\kappa= h/m$ in a Bose superfluid and $\kappa= h/2m $
in a Fermi superfluid. One can repeat the analysis of the momentum balance
for a cylindrical region around the vortex line. Now the velocity $\vec v_0$
is the superfluid velocity $\vec v_s$ far from the vortex line and the
Bernoulli law is used for variation of the chemical potential or the electric
potential in the neutral and the charged superfluids respectively. But the
momentum of the superfluid component is not conserved because of interactions
with quasiparticles (and impurities in the case of superconductors) in the
vicinity of the vortex line. One may assume that all these interactions are
incorporated by the external force $\vec F$, localized at the vortex line.
Then one has instead of Eq. (\ref{Magnus}):
\begin{equation}
\rho_s [(\vec v_L - \vec v_s) \times \vec \kappa]
=\vec F~.
   \label{Magnus-S} \end{equation}
The force $\vec F$, which enters the theory as a $\delta$-function force, is
distributed over a small vicinity of the vortex line in reality. The
dimension of this vicinity is not necessary to be of the vortex-core size,
but must be smaller than  all relevant hydrodynamic scales (e.g., the
intervortex distance, or the vortex line curvature radius).

Replacing the external force by the Magnus force, the Euler equations for the
neutral and the charged superfluids are
\begin{equation}
\frac{\partial \vec v_s}{\partial t}+ (\vec v_s \cdot \vec \nabla)\vec v_s
= - \vec \nabla \mu  + [(\vec v_L -\vec v_s) \times \vec \kappa]
\delta_2(\vec r)   ~,
   \label{Euler-s} \end{equation}
\begin{equation}
\frac{\partial \vec v_s}{\partial t}+ (\vec v_s \cdot \vec \nabla)\vec v_s =
{e \over m} \left( \vec E + {1 \over
c} [\vec v_s \times \vec B ]\right)
+ [(\vec v_L - \vec v_s) \times \vec \kappa] \delta_2(\vec r)   ~.
   \label{Euler-sup} \end{equation}

One can transform  the Euler equation using the vector identity
\begin{equation}
(\vec v_s \cdot \vec \nabla)\vec v_s = \vec \nabla {v_s^2 \over 2} - \vec v_s
\times [\vec \nabla \times \vec v_s]~.
      \label{BAC} \end{equation}
In a neutral superfluid vorticity is concentrated on the vortex line: $[\vec
\nabla \times \vec v_s] =\vec \kappa \delta(\vec r)$.  But in a
superconductor $\vec \nabla \times \vec v_s = \vec \kappa \delta(\vec r) - {e
\over mc} \vec H$.

Then the Euler equations are
\begin{equation}
\frac{\partial \vec v_s}{\partial t}= -  \vec \nabla \left( \mu +
{v_s^2 \over 2}\right) + [\vec v_L  \times \vec \kappa] \delta_2(\vec r)~,
   \label{Euler-L} \end{equation}
\begin{equation}
\frac{\partial \vec v_s}{\partial t} =
{e \over m}  \vec E -\vec \nabla \left({v_s^2 \over 2}
\right) + [\vec v_L  \times \vec \kappa] \delta_2(\vec
r)   ~.
   \label{Euler-supL} \end{equation}

This analysis demonstrates that the total external force on the superfluid in
the vicinity of the vortex line is {\em exactly } balanced by the superfluid
Magnus force $\rho_s [(\vec v_L -\vec v_s) \times \vec \kappa]$. In fact, the
term $\propto \vec v_L$ in the Euler equation may be received from a pure
kinematics: it presents the flow of the vortex lines across the line between
two points   which changes the phase difference between them (the phase
slip).  After replacing the external force by the Magnus force in the Euler
equation, the latter does not contain any information on the nature and the
magnitude of the external force. But the Euler equation is not sufficient for
description of superfluid motion: an additional equation for the vortex
velocity $\vec v_L$ is necessary. In order to derive it, one should specify
the force $\vec F$. This may be done by considering the momentum balance of
the whole liquid, but not only its superfluid component.

\subsection{Equation of vortex motion and effective Magnus force}

The momentum balance for the whole liquid in a cylindrical region around the
vortex line has been studied in Ref. \cite{S} using the collisionless kinetic
equation for quasiparticles and assuming translational invariance. The balance
yielded the equation of vortex motion which is a linear relation imposed on
three velocities $\vec v_s$, $\vec v_n$, and $\vec v_L$. Since translational
invariance was assumed, this equation depended only on velocity differences,
but not their absolute values. However, one can also include into this
balance some interactions with external fields, e.g., with impurities in
superconductors. Assuming the axial symmetry in the plane normal to the
vortex line, the momentum balance yields the following equation of vortex
motion:
\begin{equation}
\rho_s [(\vec v_L - \vec v_s) \times \kappa]
 = -D(\vec v_L - \vec v_n) - D'[\hat z \times (\vec v_L - \vec v_n)]
 -d\vec v_L - d'[\hat z \times \vec v_L]~.
     \label{vort-mot} \end{equation}
Comparing it with Eq. (\ref{Magnus-S}), one obtains the expression for $\vec
F$:
\begin{equation}
\vec F = -D(\vec v_L - \vec v_n) - D'[\hat z \times (\vec v_L - \vec v_n)]
 -d\vec v_L - d'[\hat z \times \vec v_L]~.
     \label{F} \end{equation}
The forces proportional to $D$ and $D'$ are due to scattering of free
quasiparticles by the vortex, therefore they are proportional to the
difference between the drift velocity of quasiparticles (the normal velocity
$\vec v_n$) and the vortex velocity $\vec v_L$. The forces  proportional to
$d$ and $d'$ are due to interaction between the vortex line and impurities
which are frozen into the crystal and therefore do not move if the crystal is
at rest (The case when the crystal is not at rest is discussed in Ref.
\cite{US}). These forces include also the interaction of impurities with the
quasiparticles bound in the vortex core, and therefore moving with $\vec
v_L$, but not with $\vec v_n$.

One can rewrite the equation (\ref{vort-mot}) of vortex motion collecting
together the terms proportional to the velocity $\vec v_L$:
\begin{equation}
\rho_M [\vec v_L \times \vec \kappa] + \eta \vec v_L = \rho_s [\vec v_s
\times \vec \kappa]
  +D \vec v_n  + D'[\hat z \times \vec v_n ]~.
   \label{Magnus-L} \end{equation}
Here $\rho_M= \rho_s -(D' +d')/\kappa$ and $\eta = D+d$. The forces on the
right-hand side are driving forces produced by the superfluid and normal
flows. In the theory of superconductivity the force $\vec F_L=\rho_s [\vec
v_s \times \vec \kappa] =(1/c)[\vec j_s \times \vec  \Phi_0]$, proportional
to the superfluid velocity $\vec v_s$ (or to the supercurrent $j_s=en_s
v_s$), is called the Lorentz force. Here $\Phi_0=hc/2e$ is the magnetic-flux
quantum and the vector $\vec \Phi_0$ is parallel to $\vec \kappa$. There are
also forces on the vortex produced by the normal current $\vec j_n = en_n
\vec v_n$. One can find discussion of the effect of the normal-current force
on electrodynamics of a type II superconductors in Ref. \cite{CE}.

The left-hand side of Eq. (\ref{Magnus-L}) presents the response of the
vortex to these driving forces. The factor $\rho_M$, which determines the
amplitude of the effective Magnus force on the vortex  is not equal to the
superfluid density $\rho_s$ in general: it may be more or less than
$\rho_s$.  Note that the Hall conductivity is governed by $\rho_M$, but not
by $\rho_s$. At low magnetic fields the normal current is small compared to
the supercurrent, i.e. the total current $\vec j \approx \vec j_s =e n_s \vec
v_s$ and one may neglects the terms $\propto \vec v_n$ on the right-hand side
of Eq. (\ref{Magnus-L}). On the other hand, the electric field is connected
with the vortex velocity by the Josephson relation $\vec E= {1 \over c}[\vec
H \times \vec v_L]$. Then the equation of vortex motion is equivalent to the
Ohm law connecting the current and the electric field. One can easily check
that the Hall component of the conductivity is linear in $\rho_M$.

In the superfluidity theory they usually present the equation of vortex motion
using the mutual friction parameters $B$ and $B'$ introduced by Hall and Vinen
\cite{HV}. Because of translational invariance $d=d'=0$ for superfluids, and
neglecting the normal motion ($\vec v_n=0$) the equation is
\begin{equation}
\vec v_L= \left(1- {\rho_n \over 2\rho}B' \right) \vec v_s +{\rho_n \over
2\rho} B[\hat z \times \vec v_s]= \frac{\rho_s \rho_M \kappa^2}{D^2 +
(\rho_M\kappa)^2} \vec v_s + \frac{\rho_s  \kappa D}{D^2 +
(\rho_M\kappa)^2} [\hat z \times \vec v_s]~.
       \label{HV-B} \end{equation}

In the next section we shall calculate the amplitude $D'$ of the Iordanskii
force analyzing the interaction of the vortex with phonons in the Born
approximation..

\section{Iordanskii force and Aharonov-Bohm effect} \label{Iord}

\subsection{Nonlinear Schr\"odinger equation and superfluid  hydrodynamics}
\label{Schrod}

The Gross-Pitaevskii theory \cite{GP} has suggested the nonlinear
Schr\"odinger equation to describe a weakly nonideal Bose gas:
\begin{equation}
i\hbar \frac{\partial \psi}{\partial t}= -\frac{\hbar^2}{2m} \nabla^2 \psi
+V|\psi|^2 \psi~.
   \label{Schr} \end{equation}
Here $\psi=a\exp(i\phi)$ is the condensate wave function and $V$ is the
amplitude of two-particle interaction. Using the Madelung transformation
\cite{D}, this equation for a complex function may be transformed into two
real equations for the liquid density $\rho=ma^2$ and the liquid velocity
$\vec v ={\kappa \over 2\pi} \vec \nabla \phi$ where $\kappa =h /m$ is the
circulation quantum. Far from the vortex line these equations are
hydrodynamic equations for an ideal inviscid liquid:
\begin{equation}
{\partial \rho \over \partial t} + \vec \nabla(\rho   \vec v) =0~,
     \label{m-cont} \end{equation}
\begin{equation}
{\partial \vec v \over \partial t}+ (\vec v \cdot \vec \nabla)\vec v
= - \vec \nabla \mu~.
    \label{v-Sch} \end{equation}
Here $\mu= Va^2/m$ is the chemical potential and
$c$ is the sound velocity.

Suppose that a plane sound wave  propagates in the liquid generating the phase
variation $\phi(\vec r,t)=\phi_0 \exp (i\vec k \cdot \vec r - i\omega t)$.
 Then the liquid density and velocity are functions of the time $t$ and the
position vector $\vec r$ in the plane $xy$ and can be written in the form
\begin{equation}
\rho(\vec r,t) = \rho_0 + \rho_{(1)}(\vec r,t)~,~~~~~
\vec v(\vec r,t) =\vec v_0+ \vec v_{(1)}(\vec r,t)~,
      \label{den-v} \end{equation}
where $\rho_0$ and $\vec v_0$ are the constant density and velocity in the
liquid without the sound wave, $\rho_{(1)}(\vec r,t)$ and  $\vec v_{(1)}(\vec
r,t)={\kappa \over 2\pi} \vec \nabla \phi$ are  the changes in the density and
the velocity due to the sound wave. They should be determined from the
hydrodynamic equations (\ref{m-cont}) and (\ref{v-Sch}) after their
linearization. In particular, Eq. (\ref{v-Sch}) gives the relation between
the density variation and the phase $\phi$:
\begin{equation}
\rho_{(1)}= {\rho_0 \over c^2} \mu_{(1)}= - {\rho_0 \over c^2}
{ \kappa \over 2\pi } \left\{{\partial \phi \over \partial t}+ \vec v_0 \cdot
\vec \nabla \phi(\vec r)\right\}~.
     \label{rho-phon0} \end{equation}
Substitution of this expression into Eq. (\ref{m-cont}) yields the wave
equation for a moving liquid. The sound wave has the spectrum $\omega =ck +
\vec k \cdot \vec v_0$. The sound propagation is accompanied with the
transport of mass. In the reference frame moving with the average velocity
$\vec v_0$ of the liquid the mass transport is determined by the mass
current  $\vec j^{ph}$ which is of the second order with respect to the wave
amplitude. Averaging over the wave period, one obtains:
\begin{equation}
\vec j^{ph}(\vec k)=\langle \rho_{(1)}
\vec v_{(1)} \rangle =\rho_0 \phi_0^2 {\kappa^2 k \over 8\pi^2 c}\vec k =
n(\vec p)  \vec p~.
  \label{mas-fl} \end{equation}
This expression supposes that the plane sound wave corresponds to a number
$n(\vec p)$ of phonons with the momentum $\vec p =\hbar \vec k$ and the
energy $E=\varepsilon (\vec p) +\vec p \cdot \vec v_0 $ where $\varepsilon
(\vec p) = cp $ is the energy in the reference frame moving with the liquid
velocity $\vec v_0$. Then the total mass flow is
\begin{equation}
j=\rho_0 \vec v_0 +{1 \over h^3} \int d_3\vec p n(\vec p) \vec p~.
    \label{tot-mass-fl} \end{equation}

In the thermal equilibrium at $T>0$, the
phonon numbers are given by the Planck distribution $n(\vec p)=n_0(E, \vec
v_n)$   with the drift velocity $\vec v_n$ of quasiparticles:
\begin{equation}
n_0(E, \vec v_n) =\frac{1}{\exp{E(\vec p)- \vec p \cdot \vec v_n \over T} -1}
=\frac{1}{\exp{\varepsilon(\vec p)+ \vec p \cdot (\vec v_0 -\vec v_n)
\over T} -1} ~.
   \label{BE} \end{equation}

Linearizing Eq. (\ref{BE}) with respect to the relative velocity $\vec v_0
-\vec v_n$ one can see that this expression is equivalent to the two-fluid
expression $\vec j = \rho \vec v_s + \rho_n (\vec v_n-\vec v_s) = \rho_s \vec
v_s + \rho_n \vec v_n$ assuming that $\rho=\rho_0=\rho_s + \rho_n$, $\vec v_0
= \vec v_s$, and the normal density is given by the usual two-fluid
expression:
\begin{equation}
\rho_n = -{1 \over 3h^3} \int\frac{\partial n_0(\varepsilon, 0)}{\partial
E} p^2 \, d_3 \vec p~.
     \label{rho-n} \end{equation}

This simple analysis demonstrates that two-fluid hydrodynamics with phonon
quasiparticles is identical to the nonlinear Schr\"odinger equation with
thermally excited sound waves. A next step is to analyze scattering of
phonons by the vortex in hydrodynamics of an ideal liquid.

\subsection{Scattering of phonons by the vortex in hydrodynamics} \label{Born}

The phonon scattering by a vortex line was studied beginning from the works by
Pitaevskii \cite{Pit} and Fetter \cite{Fet}. Interaction is weak and one may
use the perturbation theory (the Born approximation).

Let us consider a sound wave  $\phi(\vec r,t)=\phi_0 \exp (i\vec k \cdot \vec
r - i\omega t)$   propagating in the plane $xy$ normal to a vortex line along
the axis $z$. Then in linearized hydrodynamic equations of the previous
section the fluid velocity $\vec v_0$ should be replaced by the velocity
$\vec v_v(\vec r,t)$ around the vortex line which now depends not only on the
position vector $\vec r$, but also on time, since the vortex line is not at
rest when the sound wave propagates past the vortex. This means that $\vec r$
in Eq. (\ref{v-v}) must be replaced by $\vec r - \vec v_L  t$ and   $\partial
\vec v_v/\partial t = -(\vec v_L \cdot \vec \nabla) \vec v_v = - \vec \nabla
(\vec v_L \cdot \vec v_v)$. Since there is no external force on the liquid,
the vortex moves with the velocity in the sound wave: $\vec v_L = \vec
v_{(1)}(0,t)$. In the presence of the vortex  line linearized hydrodynamic
equations are:
\begin{equation}
{\partial \rho_{(1)} \over \partial t} + \rho_0 \vec \nabla \cdot \vec
v_{(1)} +\vec v_v \cdot \vec  \nabla \rho_{(1)} =0~,
     \label{rho} \end{equation}
\begin{equation}
{\partial \vec v_{(1)} \over \partial t}= - \vec \nabla \mu - \vec \nabla
(\vec v_v \cdot \vec v_{(1)}) - {\partial \vec v_v \over \partial t}~.
    \label{v-phon} \end{equation}
Since $\vec \nabla \mu= {c^2 \over \rho_0} \vec \nabla \rho_{(1)} $, Eq.
(\ref{v-phon}) yields:
\begin{equation}
\rho_{(1)}= - {\rho_0 \over c^2}
{ \kappa \over 2\pi } \left\{{\partial \phi \over \partial t}+ \vec v_v \cdot
[\vec \nabla \phi(\vec r) -\vec \nabla \phi(0) ]\right\}~.
     \label{rho-phon} \end{equation}
Finally the linear equation for the phonon-induced phase is
\begin{equation}
\frac{\partial^2 \phi}{\partial t^2} - c^2\vec \nabla^2\phi= - 2\vec v_v
(\vec r) \cdot \vec \nabla \frac{\partial}{\partial t} \left[ \phi(\vec r)-
\frac{1}{2} \phi(0) \right]~.
   \label{phi} \end{equation}

In the Born approximation one treats interaction with the vortex velocity
field [the right-hand side of Eq. (\ref{phi})] as a perturbation. Then
\begin{equation}
\phi =\phi_0 \exp ( - i\omega t)\left\{ \exp (i\vec k \cdot \vec r)
+{i k\over 4c} \int d_2\vec r_1 H_0^{(1)} (k|\vec r - \vec r_1|) \vec k \cdot
\vec v_v(\vec r_1) [2\exp(i\vec k \cdot \vec r_1) - 1]\right\}
        \label{phi-per} \end{equation}
Here $H_0^{(1)}(z)$ is the zero-order Hankel function of the first kind and
${i\over 4}H_0^{(1)}(k|\vec r - \vec r_1|)$ is the Green function for the 2D
wave equation. i. e. satisfies to the equation
\begin{equation}
(k^2 - \vec \nabla^2)\phi (\vec r)=\delta_2(\vec r- \vec r_1)~.
         \label{Green} \end{equation}
In our problem the small perturbation parameter is $\kappa k/c$ which is on
the order of the ratio of the wavelength $2\pi/k$ to the vortex core radius
$r_c \sim \kappa/c$.

The standard procedure in the scattering theory is the following. One uses the
asymptotic expression for the Hankel function at large values of the argument:
\begin{equation}
\lim_{z \rightarrow \infty} H_0^{(1)}(z) = \sqrt{2\over \pi z}e^{i(z-\pi/4)}~.
     \label{asym} \end{equation}
Then it is assumed that the perturbation is confined to a finite vicinity
of the line, where $r_1 \ll r$, and
\begin{equation}
|\vec r - \vec r_1| \approx r - \frac{(\vec r_1 \cdot \vec r)}{r}~.
          \label{r-r-1} \end{equation}
Finally the phase field at large values of $kr$ may be presented as a
superposition of the plane wave $\propto \exp (i\vec k \cdot \vec r)$ and the
scattered wave $\propto \exp (ik r)$:
\begin{equation}
\phi = \phi_0 \exp ( - i\omega t)\left[ \exp (i\vec k \cdot \vec r) +
\frac{ia(\varphi)}{\sqrt{r}} \exp(ikr)\right]~.
       \label{asymptot} \end{equation}
Here $a(\varphi)$ is the scattering amplitude which is a function of the angle
$\varphi$ between the initial wave vector $\vec k$ and the wave vector $\vec
k' =k\vec r/r$ after scattering. For scattering of phonons by the vortex the
Born approximation yields that \cite{Pit}
\begin{equation}
a(\varphi) =-\sqrt{{k \over 2\pi}}\frac{1}{c}e^{i{\pi \over 4}} [\hat
\kappa \times \vec k']\cdot \vec k \left({1 \over q^2} -{1\over 2k^2} \right)
= {1 \over 2}\sqrt{k\over 2\pi}\frac{\kappa }{c}e^{i{\pi \over 4}}
\frac{\sin \varphi \cos \varphi}{1-\cos \varphi}~,
     \label{a-vort} \end{equation}
where $\vec q =\vec k - \vec k'$ is the momentum transferred by the scattered
phonon to the vortex, and $q=2k\sin (\varphi/2)$. The second therm $1/2k^2$ in
parenthesis is due to the vortex line motion \cite{VorMot}.

In order to find a force on the vortex from the sound wave, one must
determine a  contribution of the wave to the average total momentum flux
$F^{ph}_i =\int dS_j \Pi^{ph}_{ij}$ over the cylindrical surface around the
vortex line where
\begin{equation}
\Pi^{ph}_{ij} =\langle P_{(2)}
\rangle \delta_{ij} + \langle \rho_{(1)} (v_{(1)})_i \rangle v_{vj} + \langle
\rho_{(1)} (v_{(1)})_j  \rangle v_{vi} + \rho_0 \langle (v_{(1)})_i
(v_{(1)})_j \rangle~.
      \label{mom-ph} \end{equation}
Here $P_{(2)}=\rho_0 \mu_{(2)} + {\partial \rho \over \partial \mu}
{\rho_{(1)}^2 \over 2}$ is the second-order phonon contribution to the
pressure where ${\partial \rho \over \partial \mu}=\rho_0/c^2$. According to
the Euler equation (\ref{Euler}) the second-order contribution  to the
chemical potential is $\mu_{(2)}=-{v_{(1)}^2 \over 2}$. Then
\begin{equation}
\langle P_{(2)} \rangle = {c^2 \over \rho_0}{\langle \rho_{(1)}^2 \rangle
\over 2} - \rho_0 {\langle v_{(1)}^2\rangle \over 2}~.
     \label{P} \end{equation}

In Appendix \ref{Asympt} it is shown that if the perturbation of the sound
wave is confined to a finite vicinity of the line, the force on the line from
the sound wave,
\begin{equation}
\vec F^{ph} =  \sigma_{\parallel} c \vec j^{ph} -
\sigma_{\perp} c [\hat z \times \vec j^{ph}]~,
       \label{force} \end{equation}
is determined by two effective cross-sections: the transport
cross-section for the dissipative force component,
\begin{equation}
\sigma_\parallel = \int \sigma (\varphi)(1-
\cos \varphi) d\varphi~,
    \label{sig-C} \end{equation}
and the transverse cross-section for the transverse force component,
\begin{equation}
\sigma_\perp = \int \sigma (\varphi) \sin \varphi d\varphi~.
    \label{sig-S} \end{equation}
The differential cross-section $\sigma (\varphi)=|a(\varphi)|^2$ in these
expressions is known due to  Eq. (\ref{a-vort}) for the scattering amplitude
$a(\varphi)$. It is quite natural that in the Born approximation the
transverse cross-section vanishes since the differential cross-section is
quadratic in the circulation $\kappa$.

However, the standard scattering-theory approach fails to describe the phonon
scattering at small angles $\varphi$. Indeed, the velocity $v_v$ induced
around the vortex is decreasing very slowly, as $1/r$. Therefore the terms
$\propto  v_v$ in the phonon momentum flux, Eq. (\ref{mom-ph}), are important
in the momentum balance. In addition, the scattering amplitude is divergent
at $\varphi \rightarrow 0$:
\begin{equation}
 \lim_{\varphi \rightarrow 0} a(\varphi) =
\sqrt{{k \over 2\pi}}\frac{\kappa }{c}e^{i{\pi \over 4}}\frac{1}{\varphi}~.
     \label{a-sm-phi} \end{equation}
This divergence is integrable in the integral for the transport cross-section,
Eq. (\ref{sig-C}). So the calculation of the transport cross-section is
reliable. Contrary to it, the integrand in Eq. (\ref{sig-S}) for the
transverse cross-section has a pole at $\varphi =0$, and the contribution of
this pole requires an additional analysis. A proper analysis of the phonon
small-angle scattering was fulfilled in Refs. \cite{I,S}. However, in the
recent publication Demircan, Ao, and Niu \cite{DAN} considered the phonon
scattering by the vortex ignoring special features of the small-angle
scattering. This is the reason why they could not find the transverse force
from phonons on the vortex.

\subsection{Small-angle phonon scattering in the Born approximation and the
Iordanskii force} \label{sm-angle}

At small scattering angles $\varphi \lesssim 1/\sqrt{kr}$ the asymptotic
expansion given by Eq. (\ref{asymptot}) does not hold. The accurate
calculation of the integral in Eq. (\ref{phi-per}) for small angles was done
in Ref. \cite{S}. A simplified version of this calculation is presented in
Appendix \ref{Interf}. It yields that at  $\varphi \ll 1$
\begin{equation}
\phi =\phi_0 \exp ( - i\omega t+i\vec k \cdot \vec r)
\left[1+\frac{i\kappa k}{2c} \Phi\left(\varphi \sqrt{kr \over
2i}\right)\right]~.
        \label{phi-small} \end{equation}
 Using an asymptotic expression for the error integral
\begin{equation}
\Phi(z)={2 \over \sqrt{\pi}} \int \limits_0^z e^{-t^2} dt
\begin{array}[t]{c}\longrightarrow \\ {z \rightarrow \infty} \end{array}
{z \over |z|} +  \sqrt{2 \over \pi z} \exp
(-z^2)
     \label{Phi} \end{equation}
at large $z$, one obtains  for angles $ 1 \gg \varphi \gg
1/\sqrt{kr}$:
\begin{equation}
\phi =\phi_0 \exp ( - i\omega t)\left[ \exp(i\vec k \cdot \vec r)
\left(1+\frac{i\kappa k}{2c} {\varphi \over |\varphi|}  \right) +
\frac{i\kappa
}{c}\sqrt{k \over 2\pi r} {1 \over \varphi} \exp\left(ikr + i{\pi \over
4}\right) \right] ~.
        \label{phi-clas} \end{equation}
The second term in brackets coincides with scattering wave at small angles
$\varphi \ll1$ when the scattering amplitude is given by Eq. (\ref{a-sm-phi}).
But now one can see that the standard scattering theory misses to reveal a
very important non-analytical correction to the incidental plane wave. We
shall see in  Sec. \ref{quasicl} that the factor $\pm \kappa k /c$ which
determines this correction is exactly the phase shift of the sound wave along
the quasiclassical trajectories past the vortex on the right and left sides.
This is a manifestation of the Aharonov-Bohm effect \cite{AB}: the sound wave
after its interaction with the vortex velocity field has different phases on
the left and on the right of the vortex line, and this phase difference
results in an interference.

In the interference region the velocity induced by the sound wave is obtained
by taking the gradient of the phase given by Eq. (\ref{phi-small}).
The velocity component normal to the wave vector $\vec k$,
\begin{equation}
v_{(1)\perp}= {\kappa \over 2\pi r} {\partial \phi \over \partial \varphi}
= \phi_0 \exp(-i\omega t + ikr)
{\kappa^2 k \over 2\pi c} \sqrt{k \over 2\pi r} ~,
       \label{v-trans} \end{equation}
determines the interference contribution to the transverse force:
\begin{equation}
\int dS_j \rho_0 \langle (v_{(1)})_\perp (v_{(1)})_j \rangle
= \int \rho_0 \langle (v_{(1)})_\perp (v_{(1)})_r \rangle rd\varphi=
{1 \over 8\pi\sqrt{\pi}} \rho_0 \phi_0^2 \frac{\kappa^3 k^2}{c}\sqrt{kr}
\int d\varphi \, \cos\left({1 \over 2} kr\varphi^2 \right) =
{1 \over 8\pi} \rho_0 \phi_0^2 \frac{\kappa^3 k^2}{c}~.
    \label{f-inter} \end{equation}
Note, that this force contribution arises from the interference region with
the transverse dimension $d_{int}\sim \sqrt{r_0 /k}$. Here $r_0$ is the large
distance from the vortex line where the momentum balance is considered. But
the interference region corresponds to very small scattering angles $\sim
d_{int}/r_0=1/\sqrt{kr_0}$. Thus  an infinitesimally small angle interval
yields a finite contribution to the transverse force. One couldn't reveal
such a contribution from the standard scattering theory using the differential
cross-section.

Exactly the same contribution to the transverse force, as in Eq.
(\ref{f-inter}), arises from the term $ v_{v_i}\langle  \rho_{(1)}
(v_{(1)})_j \rangle$ in the momentum-flux tensor, Eq. (\ref{mom-ph}). In this
term the mass flow  $\langle  \rho_{(1)}  (v_{(1)})_j \rangle$  for the plane
wave may be used [see Eq.~(\ref{mas-fl})], since we take into  account only
terms which are of the first-order in the Born parameter $\kappa k/c$.
Finally this yields the transverse force  given by the transverse
cross-section $\sigma_\perp = \kappa /c$ which is linear in the circulation
quantum  $\kappa$ and therefore cannot be obtained from the differential
cross-section quadratic in $\kappa$. This is the Iordanskii force. We shall
rederive it in the next section using expansion in partial waves in order to
demonstrate the analogy with the Aharonov-Bohm effect.

\subsection{Partial-waves analysis and the Aharonov-Bohm effect} \label{AB-PW}

Neglecting the terms of the second order in $v_v$, the equation
\begin{equation}
k^2 \phi- \left(-i\vec \nabla + {k \over c} \vec v_v\right)
^2\phi= 0
   \label{phon-vor} \end{equation}
is equivalent to Eq. (\ref{phi}) written for the harmonic sound wave with
frequency $\omega=ck$. The contribution from the vortex-line motion [the
term $\propto \phi(0)$ on the right-hand side of  Eq. (\ref{phi})] is now
neglected as unimportant for the transverse force.  Equation (\ref{phon-vor}) is
analogous to an equation which describes interaction of an electron with the
magnetic flux $\Phi$ confined to a thin tube (the Aharonov-Bohm effect
\cite{AB}):  \begin{equation} E\psi(\vec r) = {1 \over 2m}\left(-i\hbar  \vec
\nabla -{e \over c}\vec A \right)^2 \psi (\vec r)~.
      \label{AB} \end{equation}
Here $\psi$ is the electron wave function with energy $E$ and the
electromagnetic vector is connected with the magnetic flux by the relation
similar to that for the velocity $\vec v_v$ around the vortex line [Eq.
(\ref{v-v})]:
\begin{equation}
\vec A =\Phi \frac{[\hat z \times \vec r]}{2\pi r^2}~.
         \label{A} \end{equation}
Let us look for a solution of this equation as a superposition of the partial
cylindrical waves using the cylindrical system of coordinates $(r,\varphi)$:
\begin{equation}
\psi= \sum \limits_l \psi_l(r)\exp(il\varphi)~.
       \label{PW} \end{equation}
The partial-wave amplitudes should satisfy equations
\begin{equation}
\frac{d^2\psi_l}{dr^2} + {1 \over r} \frac{d\psi_l}{dr}
-\frac{(l-\gamma)^2}{r^2} \psi_l + k^2 \psi_l=0~.
   \label{AB-l} \end{equation}
Here $k$ is the wave number of the electron far from the vortex so that
$E=\hbar^2k^2 /2m$ and $\gamma = \Phi/\Phi_1$ where $\Phi_1 =hc/e$ is the
magnetic-flux quantum for one electron (two times larger than  the
magnetic-flux quantum $\Phi_0=hc/2e$  for a Cooper pair). A solution of this
equation is the Bessel function $J_{|l-\gamma|}(kr)$ with the following
asymptotics at large arguments:
\begin{equation}
J_{|l-\gamma|}(kr) \longrightarrow \sqrt{2 \over \pi kr} \cos(kr - {\pi \over
2} | l-\gamma| -{\pi \over 4})~.
      \label{asimptJ} \end{equation}
On the other hand, the expansion of the plane wave in the partial
cylindrical waves is
\begin{equation}
\exp(i\vec k \cdot \vec r) =  \exp(i k r \cos\varphi)=
\sum \limits_l J_l (kr)\exp[il(\varphi+\pi/2)] ~,
   \label{pl-wave} \end{equation}
or at large $kr$:
\begin{equation}
\exp(i\vec k \cdot \vec r) = \sqrt{2 \over \pi kr}
\sum \limits_l \cos\left(kr -{\pi \over 2}l - {\pi \over
4} \right)\exp[il(\varphi+\pi/2)] ~.
   \label{pl-wave-as} \end{equation}
In order to obtain the solution ``the incoming plane wave + the scattered
cylindrical wave'' like Eq. (\ref{asymptot}), one should determine the partial
waves $\psi_l$ from the condition that the incoming components $\propto
\exp(-ikr)$ in the plane wave and in the solution  Eq. (\ref{PW}) coincide.
This yields that
\begin{equation}
\psi_l =   \sqrt{2 \over \pi kr} \exp\left[ i{\pi \over 2} (l-|l-\gamma|)
\right]  \cos(kr - {\pi \over 2} | l-\gamma| -{\pi \over 4})~.
  \label{PW-ampl} \end{equation}
Then the solution of the Aharanov-Bohm problem takes the form of Eq.
(\ref{asymptot}) with the scattering amplitude
\begin{equation}
a(\varphi) =\sqrt{1 \over 2\pi k}\exp\left(i{\pi \over 4}\right)
\sum \limits_l [1 - \exp(2i\delta_l)] \exp(il\varphi)~,
      \label{scat-phase}  \end{equation}
where
\begin{equation}
\delta_l =(l-|l-\gamma|)\pi/2
  \label{ph-sh-AB} \end{equation}
is the partial-wave phase shift.

Equation (\ref{sig-S}) for the transverse cross-section  may be rewritten as
an expansion in partial waves \cite{Clea}:
\begin{equation}
\sigma_\perp = \int |a(\varphi)|^2 \sin \varphi d\varphi
={1 \over k}\sum \limits_l \sin2(\delta_l - \delta_{l+1})~.
    \label{sig-S-pw} \end{equation}
Using the phase shift values for the Aharonov-Bohm effect, Eq.
(\ref{ph-sh-AB}), the transverse cross-section is
\begin{equation}
\sigma_\perp=-{1 \over k}\sin 2\pi \gamma~.
         \label{sig-S-AB} \end{equation}
One can obtain  the cross-section for phonon scattering from this expression
assuming that $\gamma=-\kappa
k/2\pi c$ and expanding the sine-function in small $\gamma$.

The cross-section for the Aharonov-Bohm effect  is periodical in the magnetic
flux with the period equal to the one-electron flux quantum $\Phi_1$. If the
electron is scattered by the Cooper-pair flux quantum $\Phi_0=\Phi_1/2$ the
transverse cross-section vanishes. But presented analysis of the
Aharonov-Bohm effect is based on the assumption that the total magnetic flux
is concentrated in a very thin tube. Namely, the radius of the tube must be
much less than the electron wavelength.  This condition certainly doesn't
hold in superconductors, where the electron wavelength is on the order of the
interatomic distance  and the effective radius of the magnetic-flux tube is
the  London penetration  depth. Therefore one should consider scattering of
electrons by a thick magnetic-flux tube \cite{Clea}. Scattering of the BCS
quasiparticles by the magnetic field of the vortex is contributes to the
transverse force on the vortex, but this contribution cancels with the bulk
electromagnetic force on the superfluid electrons \cite{GS}.

In their original paper \cite{AB} Aharonov and Bohm considered only the effect
of the magnetic-flux tube on the electron wave. The force from the electrons
on the fluxon was considered later by Aharonov and Casher \cite{AC} and
therefore is called the Aharonov-Casher effect. So for our problem of
interaction between quasiparticles and the vortex both effects, Aharonov-Bohm
and Aharonov-Casher, are important. But these effects are so close one to
another that throughout the paper we use only the name ``Aharonov-Bohm
effect''.

\subsection{The transverse force in the quasiclassical theory} \label{quasicl}

Now we shall show that the Iordanskii force follows also from the
quasiclassical theory of scattering by the vortex, despite one cannot use the
quasiclassical theory for phonons. But the quasiclassical theory is valid to
describe the roton contribution to the transverse force.

Let us consider  a quasiparticle with an arbitrary spectrum $\varepsilon(p)$.
If the quasiparticle moves in the velocity field induced by the vortex, its
energy is $E(\vec p)=\varepsilon(p) + \vec p \cdot \vec v_v$ which may be
treated as a Hamiltonian to write the classic equation of motion:
\begin{eqnarray}
{d\vec r\over dt}=\frac{\partial E}{\partial \vec p} = v_G {\vec p \over p} +
\vec v_v~,\nonumber \\
{d \vec p \over dt}= - \frac{\partial E}{\partial \vec r} = - {\partial \over
\partial \vec r}(\vec p \cdot \vec v_v)~,
       \label{Hamil} \end{eqnarray}
where $v_G(p) = d \varepsilon /dp$ is the quasiparticle group velocity. As
well as in phonon scattering, we assume that the roton moves in the plane $xy$
normal to the vortex. From this equations one can find the classical
trajectory of the quasiparticle moving past the vortex line. Usually it is
close to a straight line, and a distance of the straight trajectory from the
vortex line is the impact parameter $b$.  We are looking for the transverse
force only, then we need only the variation $\delta p_\perp $ of the momentum
component normal to the initial momentum $\vec p$ which results from
quasiparticle motion past the vortex. In the first order with respect to
$\vec v_v$ Eqs. (\ref{Hamil}) yield that
\begin{equation}
\delta p_\perp(b) = - {\partial \over
\partial b}\int \limits_{-\infty}^\infty dl\,{p \over 2\pi v_G } { \kappa b
\over b^2 + l^2 }~.
       \label{p-tr} \end{equation}
where $l$ is the coordinate along the trajectory. On the
other hand, the momentum of the quasiparticle is connected to the classical
action: $\vec p = \partial S /\partial \vec r$. Then $\delta p_\perp (b) =
\partial \delta S (b)/ \partial b$ where the total variation of the classical
action along the trajectory is a function of the impact parameter $b$:
\begin{equation}
\delta S(b) =  - {p  \over 2\pi v_G } \int
\limits_{-\infty}^\infty dl\, { \kappa b \over b^2 + l^2 } ~.
       \label{S} \end{equation}
The scattering angle of the quasiparticle is $\varphi \approx - \delta
p_\perp(b)/p $ and the transverse cross-section is
\begin{equation}
\sigma_\perp =\int \limits_{-\infty}^\infty  db \sin\varphi \approx \
\int \limits_{-\infty}^\infty  db \varphi (b) = \frac{ \delta S(-\infty) -
\delta S(+\infty)}{p} = {1 \over 2\pi v_G }
\int \limits_{-\infty}^\infty  db {\partial \over \partial b}
\int \limits_{-\infty}^\infty dl\,   {\kappa b \over
b^2 + l^2} ={\kappa \over v_G}~.
    \label{cl-sigma} \end{equation}
For phonons the group velocity $v_G$ is equal to the sound velocity and Eq.
(\ref{cl-sigma}) yields the correct value of $\sigma_\perp$ even though the
quasiclassical theory is not valid for phonons: there is no well-defined
classical trajectories for phonons except for large impact parameters  $b$ at
which the scattering angle $\varphi$ is negligible. But in order to obtain
the transverse cross-section $\sigma_\perp$, one should know  only the action
variation at large impact parameters whereas the derivative $\partial \delta
S(b)/db$ of the action variation, which determines the scattering angle, is
not essential.

It is important to note that the double integral of Eq. (\ref{cl-sigma}) is
improper: its value depends on what integration is done first. The correct
procedure which was justified in Ref. \cite{S} is to integrate along the
trajectory first, and to integrate over the impact parameters afterwards.  A
way to check it is the following. We choose some finite limits in the double
integral  of Eq. (\ref{cl-sigma}) which means that the integration is
restricted by some area around the vortex line. The integral depends on the
shape of this area. For example, a circular border of the area yields
$\sigma_\perp$ by a factor 2 less than that of  Eq. (\ref{cl-sigma}). But the
full solution of the collisionless kinetic equation for the quasiparticles
made in Ref. \cite{S} showed that other terms contribute to the momentum
balance, also originating from the slow decrease of the velocity field.
Taking into account all of them, we arrive again to the expression for the
force via the cross-section given by Eq.  (\ref{cl-sigma}). The order of
integrations in  Eq.  (\ref{cl-sigma}) assumes that the integration area has
a shape of a rectangular with a long side along the quasiparticle trajectory.
For such a shape all other contributions to the transverse force exactly
cancel.

The transverse cross-section $\sigma_\perp$ determines the amplitude $ D'$ of
the transverse force on the vortex in Eqs. (\ref{F}) and (\ref{Magnus-L}). For
quasiparticles moving at an arbitrary angle to the vortex line $v_G$ in Eq,
(\ref{cl-sigma}) is the component of the group velocity in the plane normal to
the vortex line. The amplitude $D'$ of the transverse force must be determined
using the Planck distribution for phonons with the drift velocity $\vec v_n$.
This distribution should be linearized with respect to the relative velocity
$\vec v_s - \vec v_n = (\vec v_s - \vec v_L)+ (\vec v_L - \vec v_n)$. But the
velocity $\vec v_s - \vec v_L$ enters the energy $E(\vec p) =\varepsilon (\vec
p) +\vec p \cdot (\vec v_s - \vec v_L)$ of the quasiparticle in the reference
frame connected with the vortex. Using the condition of the detailed balance
one can show that the quasiparticle distribution function which depends only
on the energy cannot produce a force. Thus the only part of the distribution
function which contributes to the force is that linear in the relative drift
velocity $\vec v_n - \vec v_L$. Finally the amplitude of the transverse force
is
\begin{equation}
D'=  {1 \over 3h^3} \int\frac{\partial
n_0(\varepsilon,0)}{\partial E} p^2 \sigma_\perp v_G \, d_3 \vec p~.
     \label{D} \end{equation}
Using the cross-section $\sigma_\perp$ from Eq. (\ref{cl-sigma}). one obtains
that $D'=-\kappa \rho_n$. This means that in the translationally invariant
superfluid without scattering by impurities the amplitude $\rho_M$  of the
effective Magnus force on the vortex  in Eq.  (\ref{Magnus-L}) is equal to the
total density $\rho$, but not to the superfluid density $\rho_s$. Then
neglecting the viscous forces [$D =d=d'= 0$ in Eq.  (\ref{Magnus-L})] the
vortex moves with the center-of-mass velocity.

A rather simple and universal expression  $D'=-\kappa \rho_n$ for the
Iordanskii force amplitude  tempts to claim its universal topological origin,
since $\kappa$ in this expression is a topological charge. However, in the
next section we shall see that the expression is not universal, in fact. For
quasiparticles in a BCS superconductor with energy much exceeding the gap an
additional small factor should be put in this expression.

\section{Iordanskii force for quasiparticles in  BCS
superconductors} \label{BCS}

The wave-function of quasiparticles in the BCS theory has two components,
\begin{equation}
\psi(\vec r) = \left( \matrix{ u(\vec r) \cr v(\vec r) \cr}\right)~,
     \label{spinor} \end{equation}
which are determined from the Bogolyubov-de Gennes
equations:
\begin{equation}
-{\hbar^2 \over 2m} \left(\vec \nabla^2 + k_F^2\right) u (\vec r) +
\Delta \exp [i\theta(\vec r)]  v(\vec r) =E u(\vec r)~,
     \label{BGu} \end{equation}
\begin{equation}
{\hbar^2 \over 2m}  \left(\vec \nabla ^2 + k_F^2\right) v (\vec r) + \Delta
\exp [-i\theta(\vec r)]  u(\vec r) =E v(\vec r)~.
     \label{BGv} \end{equation}
Here $k_F$ is the Fermi wave number. We neglect the magnetic field effect
which is weak if the London penetration depth is large compared to other
relevant scales. Without the vortex the order parameter phase $\theta$ is a
constant and the equations yield the well-known BCS quasiparticle spectrum
$E=  \sqrt{\xi^2 + \Delta^2}$, where $\xi = ({\hbar^2 / 2m})(k^2 - k_F^2)$ is
the quasiparticle energy in the normal Fermi-liquid.

In Refs. \cite{GS,KK-1} the Bogolyubov-de Gennes equations for a quasiparticle
passing a vortex were solved with help of the partial-wave expansion, earlier
used also in Ref. \cite{Clea}.  Quasiparticles with the energy close to the
gap  ($\xi \ll \Delta$) behave as rotons and the transverse cross-section for
them is given by Eq. (\ref{cl-sigma}) in which the group velocity for the BCS
quasiparticles is $v_G=v_F \xi/E$. Here $v_F=\hbar k_F/m$ is the Fermi
velocity.
 In the case when the quasiparticle energy is much more than the
superconducting gap, the theory yielded that the transverse cross-section
differed from the quasiclassical result of Eq.  (\ref{cl-sigma}) by the
factor $\Delta^2/2\xi^2$. Now we rederive this result using the Born
approximation similar to that for phonons in Secs. \ref{Born} and
\ref{sm-angle}.

We use the perturbation theory with respect to the gap $\Delta$ and the
gradient of the order parameter phase $\vec \nabla \theta$. Then in the
zero-order approximation $u =u_0 \exp(i\vec k \cdot \vec r)$ and $v=0$. In
the first-order approximation the second  Bogolyubov-de Gennes equation
(\ref{BGv}) yields
\begin{equation}
v =\left\{\frac{\Delta
\exp(-i\theta)}{\xi(k) + E(k)} +\frac{\Delta \exp(-i\theta)}{[\xi(k) +
E(k)]^2}{\hbar^2 \over m}(\vec k \cdot \vec \nabla \theta)\right\} u_0
\exp(i\vec k \cdot \vec r)~.
      \label{pert-v} \end{equation}
The first term in curled brackets yields a correction to the quasiparticle
energy $\propto \Delta^2$, but does not contributes to scattering which is
determined by the order-parameter phase gradients. So we keep only the second
term proportional to $\vec \nabla \theta$. Inserting it to the first
Bogolyubov-de Gennes equation (\ref{BGu}) one obtains the following equation
for the first-order correction to the plane wave:
\begin{equation}
(\nabla^2 - k^2) u_{(1)} =  (\vec k \cdot \vec \nabla \theta)
\frac{\Delta^2}{2\xi^2} u_0 \exp(i\vec k \cdot \vec r)~.
       \label{pert-u} \end{equation}
This equation is similar to equation (\ref{phi}) for the sound wave and using
this analogy one easily obtains the expression for the
transverse cross-section:
\begin{equation}
\sigma_\perp = \frac{\Delta^2}{2\xi^2} \frac{\pi}{k_F}
 = \frac{\Delta^2}{2\xi^2} \frac{\kappa}{v_F}~.
       \label{sig-S-el} \end{equation}
Since the group velocity of quasiparticles with $E \gg \Delta$ is about the
Fermi velocity $v_F$, this expression differs from Eq. (\ref{cl-sigma}) by the
factor $\Delta^2/2\xi^2$. Thus the Iordanskii force becomes small close to
$T_c$: it decreases proportionally to $\rho_s \sim \Delta^2$, like the
superfluid Magnus force.

\section{Magnus force in the Josephson junction array} \label{JJA}

In the continuum limit, the equation of  motion for a vortex has been derived
in Ref.~\cite{4}. This derivation has not revealed any force normal to the
vortex velocity. Absence of the Magnus force suggests that the vortices move
parallel to the driving force, i.e., normally to the current, and there is no
Hall effect. Then in the limit of weak dissipation the ballistic vortex
motion is possible, which is a free vortex motion without friction and the
driving force. The Hall effect and the ballistic motion are incompatible,
since the latter is a regime with a finite electrical field and no external
current which is impossible for a finite Hall resistance. Though there have
been  experimental evidences of the ballistic vortex motion \cite{2}, one may
suspect that a more  sophisticated theory would reveal the Hall resistance,
however small. In the  present section it will be shown that the Hall effect
is {\em exactly} absent in the classical limit for the JJA. It directly
follows from the symmetry of the dynamic equations.

Let us consider a conductor in a magnetic field $\vec{H}$. When its
symmetry is not less than the three-fold (which includes a
triangular and  square lattices), the Ohm law is:
\begin{equation}
\vec{E} =\rho_L \vec{I} +  \rho_H \vec{n} \times \vec{I}~,
                  \label{11} \end{equation}
where  $\vec{n}=\vec{H}/H$, $\vec{E}$ is the electrical field,
$\rho_L$  is the longitudinal resistance and $\rho_H$ is the
Hall resistance in  the magnetic field $\vec{H}$.  Now let us
consider the transformation in which the directions of  the fields
$\vec{E}$ and $\vec{H}$ and the current $\vec{I}$ are reversed:
\begin{equation}
\vec E \rightarrow - \vec E,~~~~~~~~ \vec n\rightarrow - \vec n
(\vec H
\rightarrow - \vec H) ,~~~~~~~~ \vec I  \rightarrow -
\vec I~.
                         \label{14} \end{equation}
The Ohm law Eq. (\ref{11}) is invariant with respect to this {\em
field-current inversion} only for a system without the Hall effect
($\rho_H=0$). On the microscopical level the
field-current-inversion invariance is a direct result of the {\em
particle-hole symmetry} which was shown to forbid the Hall effect
in the Ginzburg-Landau theory (see \cite{16} and the references therein).

Next we consider the JJA with the energy
\begin{equation}
{\cal E} = \frac{1}{2} \sum_{\vec l,\vec k}\left[ Q_{\vec l}
C^{-1}_{\vec l,\vec
k} Q_{\vec k} - E_J \sin{(\phi_{\vec{l}}-\phi_{\vec{k}})} \right]~,
                   \label{0} \end{equation}
and the equations of motion
\begin{equation}
  V_{\vec{l}}=\frac{\hbar}{2e}\frac{d\phi_{\vec{l}}}{dt}~,
                      \label{15}  \end{equation}
\begin{equation}
\sum_{\vec{\mu}}  C_{\vec{l},\vec{l}+\vec{\mu}}
\frac{dV_{\vec{l}+\vec{\mu}}}{dt} -
 I_C \sum_{\vec{\mu}} \sin{(\phi_{\vec{l}}-\phi_{\vec{l}+\vec{\mu}})}+
 \sum_{\vec{\mu}} \sigma_{\vec{l},\vec{l}+\vec{\mu}}
V_{\vec{l}+\vec{\mu}}=0~.
                        \label{2} \end{equation}
Here $V_{\vec{l}}$ is the electric potential, $Q_{\vec l}=\sum_{\vec \mu}
C_{\vec{l},\vec{l}+\vec{\mu}} V_{\vec{l}+\vec{\mu}}$ is the electric charge,
and $\phi_{\vec{l}}(t)$ is the  gauge invariant phase at the node specified
by the discrete two-dimensional  position vector $\vec l$, $E_J$ is the
Josephson  coupling energy, $I_C=2eE_J/\hbar$ is the critical current,
$C_{\vec{l},\vec{n}}$  and $\sigma_{\vec{l},\vec{n}}$ are the capacity and
the  conductance matrices respectively. In the external magnetic field
$\vec{H}= \vec{\nabla} \times \vec{A}$, the gauge invariant phase
$\phi_{\vec{l}}$ is not single-valued; in fact only its difference  between
neighboring nodes is well-defined:
\begin{equation}
\phi_{\vec{l} + \vec{\mu}} - \phi_{\vec{l}} =
\varphi_{\vec{l} + \vec{\mu}} - \varphi_{\vec{l}} -
\frac{2\pi}{\Phi_0}
   \int_{(\vec{l})}^{(\vec{l} + \vec{\mu})}{\vec{A} \cdot d\vec{l}}~.
                    \label{16} \end{equation}
Here  $\varphi_{\vec{l}}$ is the canonical phase at the node
$\vec{l}$, and  the
integral over $\vec{A}$ is taken between the centers of the two
neighboring
nodes. The canonical phase is not single-valued too, but  its
circulation  along
any closed path through the nodes of JJA is  always an integer
number  of $2\pi$,
while the circulation of the gauge invariant phase may be any
number depending on
the magnetic field.

When both the external current $\vec{I}$ and the magnetic field
$\vec{H}$ are
applied to the JJA, the gauge-invariant phase can be  presented as
$\phi_{\vec{l}} =\phi_{\vec{l}}^H + \phi_{\vec{l}}^I$. Here
$\phi_{\vec{l}}^H$ is
the time-independent  phase in the equilibrium state without an
external
current, and $\phi_{\vec{l}}^I$  is the time-dependent contribution
to the phase
from the external current  $\vec{I}$. The multi-valuedness of the
phase related
to the magnetic field is present only in the static  phase
$\phi_{\vec{l}}^H$:
the time-dependent dynamical  contribution $\phi_{\vec{l}}^I$ to the
phase  is single-valued. One sees then that  the field-current
inversion [Eq. (\ref{14})] simply corresponds to the change  of signs of
all phases, and the equations of motion, Eqs.~(\ref{15})  and
(\ref{2}), are invariant with respect to this transformation. It
proves that the Hall effect does not exist in the JJA, i.e., the effective
Magnus force vanishes.

The crucial point of this very simple derivation is that one can use  the {\em
static} vortex solution $\phi_{\vec{l}}^H$ for the dynamical  problem. This
assumes that singularities of the phase distribution related to the presence
of vortices are kept at rest despite the vortices  themselves are  driven by
the Lorentz force. For continuous superconductors this approach is invalid
and our derivation does not work. So there is a fundamental difference between
vortices in a lattice and vortices in a continuous superconductor. Indeed, in
the lattice there are no singular vortex lines.  They appear only in the
continuum limit. At best, one can define the lattice cell containing the
vortex center. This definition has been borrowed from the continuous theory:
it is the cell, around which the circulation of the phase   $\varphi$ is
equal to $2\pi$. However, in the lattice the circulation around a  closed
path  is not well-defined. Let us consider some closed path through a
discrete  number of nodes with the phase circulation $2\pi$. One may  change
the phase  difference by $-2\pi$ between any two neighboring nodes  on the
path without any effect  on observed physical parameters (currents,  voltages
and so on). Then the  circulation vanishes along the path considered, but
must appear along a path over other nodes. Thus one cannot locate the
position of the phase singularity.  In order to avoid this ambiguity  in the
JJA model, a  special rule has been formulated: the phase difference  between
two neighboring nodes must not  exceed $\pi$. When for some bond the phase
difference achieves the value $\pi$, one must  redefine the  phases; as a
result, the vortex center is put into another cell. This  procedure is usual
for numerical studies of the vortex motion in JJA \cite{7}. However, this
rule is not obligatory for the dynamic theory of JJA. Instead, one may keep
$2\pi$ circulations of the phase $\varphi$ at  fixed  cells during the dynamic
process without worrying where the vortex center  (defined according to the
aforementioned rule) is really located.

It is important to stress, that one cannot derive the effective Magnus force
in JJA using the continuous approach. Let us discuss this in more details. In
the continuum limit the set of discrete vectors $\vec{l}$ is  replaced with
the continuum space of $\vec{l}$. One can define the field of the canonical
(but not gauge invariant!) phase $\varphi(\vec{l})$ in this space everywhere
except for the  singular points which are the centers of the vortices with
the phase circulation  $2\pi$ around them. It is assumed that the spatial
variation of the phases is small and the phase  differences can be replaced
by the phase gradients according to
\begin{equation}
\varphi_{\vec{l}+\vec{\mu}} -\varphi_{\vec{l}} \approx (\vec{\mu}
\cdot \vec{
\nabla})\varphi(\vec{l}) \ll 1~.
                     \label{3} \end{equation}
Then one can  derive the partial differential equations for a continuous field
of the canonical phase $\varphi(\vec{l},t)$ which correspond to some
Lagrangian $L\{\varphi(\vec{l},t)\}$ (the dissipation is neglected now). The
Lagrangian may include the term proportional to the time derivative of
$\varphi(\vec{l},t)$ which is called the Wess-Zumino term \cite{Ga}:
\begin{equation}
L\{\varphi(\vec{l},t)\}={1 \over 2}q {\partial\varphi \over \partial t}
+L_0\{\varphi(\vec{l},t)\}~.
 \label{WZ} \end{equation}
It is possible to derive the equation of vortex motion from this Lagrangian
following  Refs.~\cite{4,5}. One must use the phase field for a slowly  moving
vortex:
\begin{equation}
\varphi^V(\vec{l}, t)=\arctan\frac{l_y-
y(t)}{l_x- x(t)}~,
                   \label{5} \end{equation}
where $\vec{r}(t)=[x(t),y(t)]$ is the 2D position vector of the vortex
center. Substituting the vortex solution into the  Lagrangian density given
by Eq. (\ref{WZ}) and integrating over the $xy$  plane, one obtains  the
effective Lagrangian which is now a functional of the trajectory for a moving
vortex:
\begin{equation}
    {\cal L}^V\{\vec r(t)\}=- \pi q \dot{ \vec r} \cdot [\hat{z} \times {\vec
r}]+{\cal L}_0^V\{\vec r(t)\}~.
                      \label{6} \end{equation}
Varying this Lagrangian with respect to $\vec r(t)$, one obtains the  equation
of vortex motion with the effective Magnus force $\propto q$:
\begin{equation}
 2\pi q[\dot{\vec{r}} \times  \hat{z}]
 = \vec{ F}_\Sigma~.
                       \label{10} \end{equation}
Here $\vec{ F}_\Sigma$ includes all other forces (the Lorentz force and the
inertia force) obtained from the Lagrangian ${\cal L}_0^V\{\vec r(t)\}$
without the Wess-Zumino term. However, the factor $q$ is not defined. If $q$
is constant, it has  no effect on the field equation for $\varphi(\vec l,t)$
since the Wess-Zumino term is a full time derivative in the field Lagrangian,
Eq. (\ref{WZ}). The unknown factor $q$ should be proportional to some electric
charge, since the charge is a variable conjugate to the canonical phase
$\varphi$. But it remains unclear what is this charge: either the background
charge  determined by the whole Fermi-see of the superconducting island, or
an  external charge induced outside as suggested in Ref. \cite{11.1}. Thus in
the continuum limit the problem of the effective Magnus force and the Hall
effect in the JJA remains unresolved. It must not be a surprise since in the
continuum limit the JJA model becomes translationally invariant and
``forgets'' that originally it had been a lattice model without translational
invariance. Meanwhile, the latter is crucial for the amplitude of the
effective Magnus force. An additional physical principle beyond the
continuum  theory should be  involved to obtain the equation for the vortex
velocity. This principle is provided by our symmetry analysis. According to
it $q=0$ and one should not include the Wess-Zumino term into the field
Lagrangian.

The presence of the external charge has no effect on our symmetry analysis. In
order to take into account the external charge, one should use the Gibbs
potential $G={\cal E} - V^{ex} \sum_{\vec l} Q_{\vec l} $, where ${\cal E}$ is
given by Eq. (\ref{0}) and $V^{ex}$ is the electric potential which creates
the external charge $Q^{ex}= V^{ex} \sum_{\vec \mu} C_{\vec l + \vec \mu}$.
Then introducing the charge deviation $Q'_{\vec l}=Q_{\vec l} -  Q^{ex}$, one
returns back to the energy ${\cal E}$ with $Q'_{\vec l}$ instead of $Q_{\vec
l}$. These arguments show that the external charge cannot lead to the Hall
effect: its effect is restricted with the shift of the Fermi level, but the
particle-hole symmetry is restored with respect to the new Fermi level.
However, the external electric charge may produce the Magnus force in the
quantum theory of JJA which takes into account the electron charge
quantization \cite{qJJA}. Then the Magnus force and the Hall conductivity are
periodic in the electron charge.

\section{Summary  and discussion} \label{DS}

We have shown how the Magnus force appears in the equation of motion for a
superfluid component (the superfluid Magnus force) and the equation of motion
for a vortex (the effective Magnus force). Whereas the superfluid Magnus
force proportional to the superfluid density is known exactly (from classical
hydrodynamics, or from the Berry phase approach), there is no general
expression for the effective Magnus force: it depends on interaction of the
vortex with quasiparticles and with the external fields, like those from
impurities in a dirty superconductors. Meanwhile, it is mostly the effective
Magnus force which determines the observable effects: the mutual friction in
superfluids, the Hall effect and the acoustic Faraday effect in
superconductors,  vortex quantum tunnelling.

We have calculated the contribution of quasiparticles to the effective Magnus
force for phonons in a superfluid and for high-energy quasiparticles in a BCS
superconductor using the Born approximation. The transverse force from
quasiparticles on the vortex (the Iordanskii force) originates from
interference between quasiparticles passing on different sides of the vortex
(the Aharonov-Bohm effect).

Our symmetry analysis of the classical Josephson junction array has
demonstrated that the effective Magnus force exactly  vanishes and there is no
Hall effect despite the finite superfluid density. One may formally interpret
this result that  the force from external fields violating translational
invariance exactly compensates the superfluid Magnus force, though the
analysis is not able to reveal these two forces separately, but only their
joint outcome, namely, the effective Magnus force.

The Ao-Thouless theory yields only the superfluid Magnus force which appears
in the momentum balance of the superfluid component (the condensate). Indeed,
Gaitan \cite{Ga} derived the Ao-Thouless result for a charged superfluid,
analyzing the momentum balance for the condensate. In order to derive the
effective Magnus force (the total transverse force on the vortex), one must
consider the momentum balance for the whole system.

Ao, Niu, and Thouless \cite{ANT} stated that the Iordanskii force did not
appear in their Berry phase approach. But the recent paper on the Born
quasiparticle scattering by Demircan, Ao, and Niu \cite{DAN} demonstrated that
they admitted the force from quasiparticle scattering in their approach, but
concluded that this didn't yield any transverse force on the vortex. This
conclusion was based on a wrong analysis of the Born phonon scattering missing
the contribution from the  Aharonov-Bohm interference. Thus the source of
controversy is not in a difference of approaches, but in the problem how to
calculate integrals for the Born phonon scattering.

The Ao-Thouless theory rejects also any force on the vortex from the external
fields, like the Kopnin-Kravtsov force in a dirty superconductor. On the basis
of this theory  Gaitan and Shenoy \cite{GaS} predicted the finite effective
Magnus force and the Hall effect for the Josephson-junction array. This
prediction contradicts to our symmetry analysis and to the experiment.  Gaitan
and Shenoy \cite{GaS} used in their analysis the  Wess-Zumino term in the
Lagrangian for the continuum limit of JJA. We have shown in Sec. \ref{JJA} why
this approach is not reliable. On the other hand, Zhu, Tan, and Ao \cite{ZTA}
try to conciliate the negligible Hall effect in JJA with the Ao-Thouless
theory supposing that the vortices cross  JJA  between superconducting
islands where the superfluid order parameter vanishes and therefore the
Magnus force is negligible. But the superfluid order parameter {\em does not
vanish} between the islands even though it is much smaller than in the
islands. In fact, this small order parameter determines a finite superfluid
density in the continuum limit. Thus in JJA  the Hall effect vanishes {\em
despite} a finite superfluid density, contrary to the prediction of the
Ao-Thouless theory.

Makhlin and Volovik \cite{MV} suggested that the superfluid Magnus force in
JJA is nearly compensated by the force from the bound states in the junctions
(the spectral flow of bound states). But they  did not conclude that the
compensation is complete, and assumed the Fermi superfluid in islands and the
SNS Josephson junctions. Our analysis shows that the Magnus force {\em
exactly} vanishes for JJA independently on microscopic nature of the
superconducting islands and the junctions. This shows that the core bound
states and the spectral flow are not the only explanation for compensation of
the Magnus force in the systems without translational invariance.

\vspace{.5 cm}
\centerline{\bf Acknowledgements}
\vspace{.5 cm}

I appreciate very much discussions with G. Blatter, U. Eckern, V.B.
Geshkenbein, B. Horovitz, N.B. Kopnin, G. Sch\"on, A.L. Shelankov, A. van
Otterlo, W.F. Vinen, and G.E. Volovik. The exchange of messages with D.J.
Thouless and P. Ao was very useful for better understanding of their point of
view on the problem. The work was partially supported by the Russian Fund of
Basic Researches (grant 96-02-16943-a).

\begin{appendix}
\renewcommand{\theequation}{\thesection.\arabic{equation}}
\section{The force on the line scattering the sound wave} \label{Asympt}
\setcounter{equation}{0}

First we derive the analogue of the optical theorem for the sound wave. For
the latter we use the asymptotic representation, Eq.~(\ref{asymptot}), in
which the scattering amplitude $a(\varphi)$ is not necessarily obtained in
the Born approximation. But in general $a(\varphi)$ should satisfy the
condition that the total mass flow through the cylindrical surface
surrounding the scattering line vanishes.

An asymptotic expression for the average mass flow from the sound wave is
\begin{equation}
\vec j^{ph} =\langle \rho_{(1)}\vec v_{(1)} \rangle =\rho_0
\phi_0^2 {\kappa^2 k \over 8\pi^2 c}
\left\{\vec k + \frac{|a|^2}{r}\vec k'-(\vec k + \vec k')
 \frac{1}{\sqrt{r}} [ \mbox{Im}\{ a\}\, \cos(kr - \vec
k \cdot \vec r) ]+ \mbox{Re} \{a\}\, \sin (kr - \vec
k \cdot \vec r) ] \right\}~.
      \label{as-flow} \end{equation}

The condition that the total flow through the cylindrical surface around the
scattering line vanishes  is
\begin{equation}
\int  j^{ph}_i dS_i = \int \langle \rho_{(1)} \vec v_{(1)i} \rangle dS_i =
\rho_0 \phi_0^2 {\kappa^2 k^2 \over 8\pi^2 c}\int \left\{ \cos \varphi +
{|a(\varphi)|^2 \over r} - {\mbox{Im} \{a(\varphi)\} \over \sqrt{r}} (1 +\cos
\varphi) \cos [kr(1 -\cos \varphi) ]\right\}\,d\varphi =0 ~.
        \label{mas-tot} \end{equation}
The integral over the term $\propto \mbox{Im}\{ a(\varphi)\}$ expands only
over the region of small angles since $kr \gg 1$. Finally this condition
yields
\begin{equation}
-2\sqrt{\pi \over k} \mbox{Im} \{a(0)\}+ \int |a(\varphi)|^2 d\varphi =0~.
    \label{opt-t} \end{equation}

Next let us consider the momentum balance which determines the force on the
scattering line from the sound wave $F^{ph}_i =-\int dS_j \Pi^{ph}_{ij}$
where:
\begin{equation}
\Pi^{ph}_{ij} =\left({c^2 \over \rho_0}{\langle \rho_{(1)}^2 \rangle \over 2}
- \rho_0 {\langle v_{(1)}^2\rangle \over 2} \right) \delta_{ij}
+  \rho_0 \langle (v_{(1)})_i (v_{(1)})_j \rangle~.
      \label{mom-ph-sh} \end{equation}
The pressure term vanishes after averaging, but the convection term is
essential and yields for the force on the vortex:
\begin{eqnarray}
\vec F^{ph} =-\rho_0 \phi_0^2 {\kappa^2 k \over 8\pi^2} \int
\left\{\vec k \cos \varphi  + {|a(\varphi)|^2 \over r^2}k\vec r -
{\mbox{Im} \{a(\varphi)\} \over \sqrt{r}} \cos [kr(1 -\cos \varphi) ]
\left(\vec k + k{\vec r \over r} \right) \right\}\,rd\varphi \nonumber\\
\approx
-\rho_0 \phi_0^2 {\kappa^2 k \over 8\pi^2} \int
\left[ \vec k \cos \varphi + {|a(\varphi)|^2 \over r}k\vec r  -
{\mbox{Im} \{a(0)\}\over \sqrt{r^2}} \cos \left({1 \over 2}kr \varphi^2
\right) 2\vec k  \right]\,rd\varphi
\nonumber \\
=-\rho_0 \phi_0^2 {\kappa^2 k \over 8\pi^2}
\left[ \int {|a(\varphi)|^2 \over r}k\vec r \,d\varphi - 2\sqrt{\pi \over k}
\mbox{Im} \{a(0)\} \vec k\right] ~.
        \label{force-sh} \end{eqnarray}
With help of the optical theorem Eq. (\ref{opt-t}) one obtains the expression
Eq. (\ref{force}) with the effective cross-sections determined by Eqs.
(\ref{sig-C}) and (\ref{sig-S}).

\section{Small-angle scattered sound wave} \label{Interf}
\setcounter{equation}{0}

Using the asymptotics of the Hankel function, Eq. (\ref{phi-per}) can be
rewritten as
\begin{equation}
\phi =\phi_0 \exp ( - i\omega t)\left\{ \exp (i\vec k \cdot \vec r)
+{\kappa k\over c}\sqrt{i \over 2\pi r} \int d_2\vec r_1
\exp(i\vec k \cdot \vec r_1    +ik |\vec r - \vec r_1|)
\frac{\vec k \cdot [\hat z	 \times \vec r_1]}{r_1^2} \right\}~.
        \label{phi-sm} \end{equation}
Here the effect of the vortex-line motion was neglected as irrelevant for
small-angle scattering. Expansion Eq. (\ref{r-r-1}) is not accurate enough and
next terms of the expansion must be kept:
\begin{equation}
|\vec r - \vec r_1| \approx r - \frac{(\vec r_1 \cdot \vec r)}{r}
+ \frac{ r_1^2}{2r} - \frac{(\vec r_1 \cdot \vec r)^2}{2r^3} ~.
          \label{r-r-2} \end{equation}
The terms of the second order in $r_1$ are important since the perturbation is
not well localized near the vortex line, but decreasing slowly when $r_1$ is
increasing. Using the Cartesian coordinates of the position vector $\vec
r_1(x,y)$ and the inequality $\varphi \ll 1$, one obtains
\begin{equation}
\phi =\phi_0 \exp ( - i\omega t)\left\{ \exp (i\vec k \cdot \vec r)
+{\kappa k^2\over c}\sqrt{i \over 2\pi r} \int \int dx\,dy\,
\exp\left[i k \left( r -y\varphi +{y^2 \over 2r}  \right)\right]
\frac{y}{x^2+y^2}\right\}~.
        \label{phi-sm-C} \end{equation}
The double integral in this expression may be transformed into the error
integral:
\begin{eqnarray}
\int \limits_{-\infty}^{\infty} dx \int \limits_{-\infty}^{\infty} dy\,
\exp\left[i k \left( r -y\varphi +{y^2 \over 2r}  \right)\right]
\frac{y}{x^2+y^2} \nonumber \\ = \exp\left[ikr\left(1-{\varphi^2 \over 2}
\right)\right]
 \int \limits_{-\infty}^{\infty} dy\, \pi\frac{y}{|y|}\,
\exp\left[{ik \over 2r} \left(r\varphi -y\right)^2 \right]
=- \pi\sqrt{2\pi i r
\over k}\exp\left[ikr\left(1-{\varphi^2 \over 2}\right) \right] \Phi
\left(\varphi \sqrt{kr \over 2i}\right)~.
          \nonumber \end{eqnarray}
Then Eq. (\ref{phi-sm-C}) coinsides with (\ref{phi-small}).

\end{appendix}

\end{document}